\documentclass[10pt,letterpaper,twocolumn]{article} 

\usepackage{ol2}
\usepackage[draft]{hyperref}
\usepackage{amsmath}

\begin{document}

\twocolumn[ 

\title{Airy beams from a microchip laser}


\author{Stefano Longhi}

\address{Dipartimento di Fisica, Politecnico di Milano, Piazza L. da Vinci 32, I-20133 Milano, Italy}

\begin{abstract}
It is theoretically shown that an end-pumped microchip laser formed
by a thin laser crystal with plane-plane but slightly tilted facets
can emit, under appropriate pumping conditions and near a crystal
edge, a truncated self-accelerating Airy output beam.
\end{abstract}


%

 ] 

\noindent Microchip lasers are an important class of monolithic
solid-state lasers which have attracted a considerable interest as
single-frequency and ultracompact devices \cite{m1,m2,m3}. A
microchip laser generally consists of a thin rare-earth-doped laser
crystal with dielectric coatings on both faces to form a plane
parallel cavity, end-pumped by a diode laser. Owing to the planar
nature of the cavity, the transverse modes of a microchip laser are
not determined by mirror curvatures, rather by weaker, residual
guiding effects, such as gain or thermally-induced guiding
\cite{m4,m5,m6,m7,m8,m9}.\\
In this Letter we consider a microchip laser with two planar but
slightly tilted facets, and show that end-pumping near the edge of
the laser crystal can lead to laser emission into self-accelerating
Airy beams, a class of optical beams which has received a great
interest in the recent few years \cite{a1,a2,a3,a4,a5,a6,a7,a8}.
Airy beams show rather unique properties, such as diffraction free
propagation along a curved (parabolic) path and self healing, i.e.,
restoring their canonical form after passing small obstacles. Airy
beams found recently important applications in optics, including
optical trapping \cite{a3}, plasma waveguiding \cite{a6}  and
nonlinear
frequency generation \cite{a7,a8}.\\
Let us consider a microchip laser operating at the wavelength
$\lambda$ made of a thin crystal (length $l$, refractive index
$n_s$) with two highly-reflecting and dielectric-coated planar
facets [Fig.1(a)]. For the sake of simplicity, we first consider the
case of a one transverse spatial coordinate. The two crystal facets
are not perfectly parallel each other, but are tilted to form a
small angle $\alpha_x$. The microchip is end-pumped by a Gaussian
beam, focused at a distance $\rho_x$ from the upper edge of the
crystal [see Fig.1(a)]. Field propagation in the cavity can be
described within an effective-index and mean-field model \cite{m9},
in which the intracavity field envelope $\psi$ is assumed to
slightly vary in one cavity round trip. Propagation back and forth
between the two mirrors is equivalent to propagation along a
lensguide with a complex refractive index $n(x)=n_s+\Delta n(x)
\simeq n_s$, where the real and imaginary parts of $\Delta n(x)$
account for (weak) index and gain guiding effects, respectively
\cite{m9}. Let $\psi(x)$ be the field envelope at the reference
plane $z=0$ in the cavity; after a cavity round-trip, the field
envelope is transformed into the one $\psi'(x)=\hat{T} \psi(x)$,
where $\hat{T}$ is the round-trip propagation operator given by
\begin{equation}
\hat{T}=\exp(-\gamma) \exp(-ikl \hat{H}) \exp (-ikn_s \alpha_x x)
\exp(-ikl \hat{H}). \; \;\;\;\;
\end{equation}
\begin{figure}[htb]
\centerline{\includegraphics[width=8.2cm]{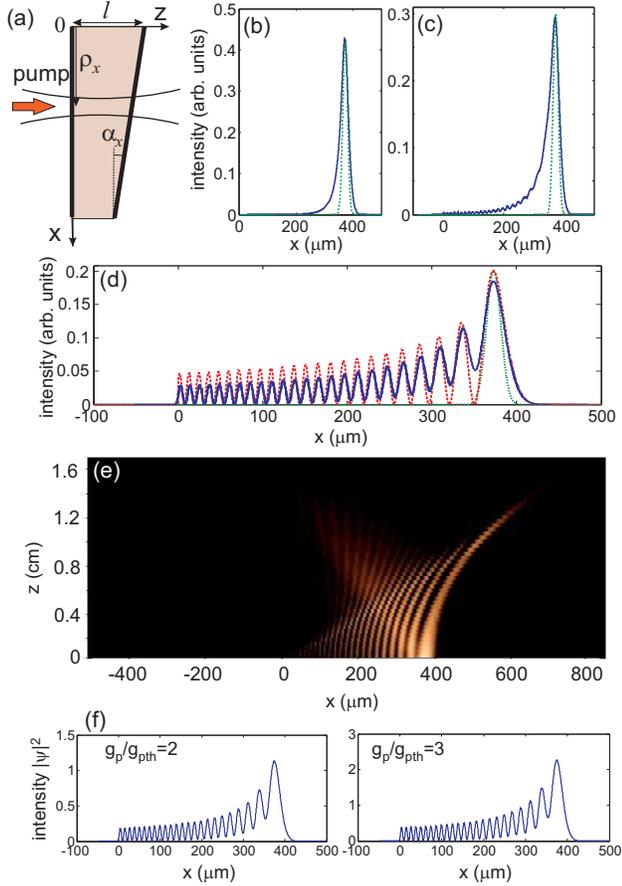}} \caption{ (Color
online) (a) Schematic of an end-pumped microchip laser with tilted
mirror. (b)-(d) Numerically-computed transverse mode profile at
lasing threshold (solid curve)
 for decreasing values of the logarithmic cavity loss, showing the transition from an asymmetric Gaussian beam to a truncated Airy beam:
(b) $\gamma=0.142$, (c) $\gamma=0.0651$, and (d) $\gamma=0.019$. The
gain threshold $g_p=g_{pth}$ is $g_{pth}=0.326$ in (b),
$g_{pth}=0.2172$ in (c), and $g_{pth}=0.1086$ in (d).
 The dotted curve in the figures is the profile of the
 Gaussian population inversion $\Delta N_0$, whereas the dashed curve in (d) is the intensity profile of the Airy mode $|\psi_n|^2$ with
$\xi_n=23.224$. (e) Free-space propagation of the output laser beam
of Fig.1(d) [intensity distribution in the $(x,z)$ plane]. (f) Same
as (d), but for a gain parameter $g_p$ two times (left panel) and
three times (right panel) above threshold.}
\end{figure}
In Eq.(1), $k=2 \pi / \lambda$ is the wave number, $\gamma$ is the
single-pass logarithmic loss of the cavity \cite{Svelto}, and
\begin{equation}
\hat{H}=-1/(2n_sk^2) \partial^2_x-\Delta n(x).
\end{equation}
Note that $\psi'(x)$ is obtained from $\psi(x)$ by application of
the four operators $\hat{T}_1=\exp(-\gamma)$,
$\hat{T}_2=\exp(-ikl\hat{H})$, $\hat{T}_3=\exp(-ik n_s \alpha_x x)$
and $\hat{T}_4=\hat{T}_2$ describing the effects of cavity loss,
forward propagation in the crystal, phase shift introduced by the
mirror tilt, and backward propagation in the crystal, respectively.
In the mean-field limit, one can assume $\hat{T} \simeq
1-\gamma-2ikl\hat{H}-ikn_s \alpha_x x$. Indicating by
$T_R=2ln_s/c_0$ the round-trip transit time in the cavity, where
$c_0$ is the speed of light in vacuum, the temporal evolution of the
field envelope $\psi(x,t)$ is then given by $\partial_t \psi(x,t)
\simeq [\psi'(x)-\psi(x)]/T_R$, i.e.
\begin{equation}
\partial_t \psi =-\frac{\psi}{2
\tau_c}+i\frac{c_0}{2n_s^2k}\partial^2_x \psi+ikc_0
\left(\frac{\Delta n}{n_s}-\frac{\alpha_x x}{2l} \right) \psi,
\end{equation}
where $\tau_c=T_R/(2 \gamma)$ is the cavity photon lifetime.
Neglecting thermal lensing effects \cite{m7} and assuming that the
laser operates at the resonance frequency of the gain medium, so
that gain-related index guiding effects can be neglected as well
\cite{m7,m9}, the refractive index $\Delta n$ in Eq.(3) can be taken
to be purely imaginary and given by $\Delta n(x)=-i \sigma_g
N(x)/(2k)$, where $\sigma_g$ is the cross-section of the lasing
transition and $N(x)$ is the (saturated) population inversion
distribution in the crystal. For a Gaussian pump beam with a
Rayleigh range larger than the crystal thickness $l$ and for the
laser operated near threshold, for which saturation effects can be
neglected \cite{m7}, one has  $N(x)=N_0(x)=N_p \exp[-2
(x-\rho_x)^2/w_p^2]$, where $w_p$ is the pump beam spot size. Above
threshold, in the mean-field limit and neglecting spatial hole
burning, following Ref.\cite{m8} saturation effects can be included
in the model by taking $N(x)=N_0(x)/(1+ |\psi|^2)$, where the
intracavity field intensity is normalized to the saturation
intensity of the lasing transition. The on-axis single-pass
(unsaturated) photon gain coefficient $g_p$ in the medium is related
to the (unsaturated) peak population inversion $N_p$ by the simple
relation $g_p=\sigma_gN_pl$ \cite{Svelto}. Crystal truncation at
$x=0$ introduces an effective barrier potential for the paraxial
lasing field because of total internal reflection, which is modeled
by requiring $\psi=0$ for $x \leq 0$ in Eq.(3). As discussed below,
such an edge effect is of major importance for the laser to
oscillate on a (truncated) Airy transverse mode. Introducing the
normalized spatial and temporal variables $\xi=x/L$ and $\tau=t/ T$,
where
\begin{equation}
L=\left( \frac{l}{n_s^2 \alpha_x k^2} \right)^{1/3} \; , \;\;\;
T=\frac{2}{c_0} \left( \frac{l^2n_s^2}{\alpha_x^2k} \right)^{1/3},
\end{equation}
 Eq.(3) can be cast in the
dimensionless form
\begin{equation}
\partial_{\tau} \psi=i \partial^{2}_{\xi} \psi- i\xi
\psi+\frac{T}{T_R} \left( \frac{g_p F}{1+|\psi|^2}-\gamma\right)
\psi \equiv \hat{Q} \psi,
\end{equation}
where $F=F(\xi)=\exp[-2(\xi L-\rho_x)^2/w_p^2]$ is the normalized
profile of the unsaturated population inversion in the crystal and
$g_p=\sigma_gN_pl$ the peak gain coefficient. The thresholds and
profiles of the various transverse modes sustained by the microchip
laser can be computed by numerical analysis of eigenvalues and
eigenfunctions of the (linear) operator $\hat{T}$ by taking $\Delta
N(x)=\Delta N_0(x)$. In particular, the lowest-order lasing mode can
be computed by a standard Fox-Li iterative method \cite{FoxLie}. The
profile of the lasing mode turns out to depend mainly on the two
parameters $(T/T_R)\gamma$ and $\rho_x/L$, i.e. by the logarithmic
cavity loss, tilting angle, and distance of the pump beam from the
crystal edge. This is because two different physical mechanisms
compete in the formation of the transverse laser mode. On the one
hand, gain guiding provided by the pump tends to localize the lasing
mode near the gain region, as in usual microchip lasers with
parallel facets \cite{m4,m5,m6,m7,m8}; on the other hand, the mirror
tilting pushes the light beam toward the edge of the crystal, where
it is back reflected owing to total internal reflection, thus
providing an effective trapping of light which was discussed in
\cite{DellaValle09} in another optical setting. Depending on the
strength of these two competing processes, two distinct transverse
confining regimes can be found. When $\rho/L_x$ is very large (i.e.
the crystal is pumped far from the crystal edge) and the cavity
losses are not too small (i.e. the gain guide is strong enough), the
gain-guiding mechanism prevails and the lasing mode looks like a
slightly asymmetric Gaussian beam, similar to the off-axis Gaussian
modes with complex arguments studied in \cite{note1,note2}.
Conversely, if the crystal is pumped closer to the edge of the
crystal and the logarithmic cavity loss $\gamma$ is sufficiently
small, the edge trapping mechanism prevails. In this case, gain and
loss terms entering on the right hand side of Eq.(5) can be treated
as perturbation terms, and the transverse mode profiles of the cold
cavity are defined by the Airy equation $\partial^2_{\xi} \psi-\xi
\psi= \sigma \psi$ with the boundary condition $\psi(0)=0$, i.e.
$\psi_n(\xi)={\rm Ai}(\xi+\xi_n)$, where $\xi_n$ is the $n$th zero
of the Airy function and $\sigma=\xi_n$ is the eigenvalue that
determines the transverse mode frequency. Hence, the lasing mode
will be determined by the (truncated) Airy eigenmode $\psi_n(\xi)$
with the largest overlap with the pump beam profile. As an example,
Figs.1(b-d) show the transition of the lasing mode at threshold from
an asymmetric Gaussian-like beam to a truncated Airy beam. In the
simulations, different crystals were considered, with decreasing
values of cavity loss $\gamma$ (controlled e.g. by the mirror
reflectance) and equal distance $\rho_x$ of the pump beam from the
crystal edge. The lowest-order (lasing) mode at threshold has been
computed by the Fox-Li iterative method using the exact round-trip
propagator $\hat{T}$, defined by Eq.(1). Parameter values used in
the numerical simulations refer to a Nd:YVO$_4$ microchip laser
\cite{m7,m8} ($n_s=2.17$, $\lambda=1064$ nm) with crystal thickness
$l=400 \; \mu$m and tilt angle $\alpha_x=0.5\; {\rm mrad}$, which
correspond to spatial and temporal scales $L \simeq 17 \; \mu$m and
$T \simeq 53$ ps [see Eq.(4)]. The pump distance from the crystal
edge is $\rho_x= 373 \; \mu$m, whereas the pump spot size is $w_p=
17 \; \mu$m. As for large cavity losses the gain guiding is the
dominant mechanism of transverse confinement and the lasing mode
looks like a slightly asymmetric Gaussian beam [Fig.1(b)], at lower
cavity losses index guiding provided by edge reflection and mirror
tilt prevails and the lasing mode is well approximated by a
truncated Airy beam with its largest lobe overlapped with the pump
beam [see Fig.1(d)]. The free-space propagation of the output beam
in this case is depicted in Fig.1(e), clearly showing the
characteristic parabolic (accelerated) path of the Airy beams. It
should be noted that the transition from Gaussian-like to Airy-like
modes shown in Fig.1 could be observed by assuming different
microchip lasers with equal loss $\gamma$ but with increasing tilt
angle $\alpha_x$. Laser emission in the Airy-like mode persists for
the microchip operated well above threshold, where saturation
effects can not be neglected. As an example, Fig.1(f) shows the
numerically-computed steady-state lasing modes, which account for
saturation of the population inversion, for the same conditions of
Fig.1(d) but for the laser operated with a gain parameter $g_p$ two
and three times above threshold. As one can see, gain saturation
does not appreciably modify the transverse  mode profile.\\
\begin{figure}[htb]
\centerline{\includegraphics[width=8.3cm]{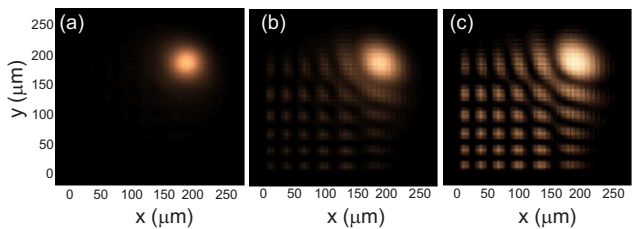}} \caption{ (Color
online) Two-dimensional transverse intensity profile $|\psi(x,y)|^2$
of the lasing mode at threshold for
 (a)
$\gamma=0.0632$, (b) $\gamma=0.0298$, and (c) $\gamma=0.0106$ (other
parameter values are given in the text). The gain threshold is
$g_{pth}=0.354$ in (a), $g_{pth}=0.236$ in (b), and $g_{pth}=0.118$
in (c).}
\end{figure}
The previous analysis can be extended to the case of two-transverse
spatial variables $x$ and $y$. Indicating by $\alpha_x$ and
$\alpha_y$ the mirror tilting angles with respect to the $x$ and $y$
axes, in terms of the dimensionless variables $\xi=x/L_x$,
$\eta=y/L$ and $\tau=t/T$, where $L$ and $T$ are defined by Eq.(4),
the evolution equation of the field envelope $\psi$ in the
mean-field limit reads $\partial_{\tau} \psi=\hat{Q} \psi$, where
$\hat{Q}=i [\partial_{\xi}^2+\partial_{\eta}^2-\xi -(\alpha_y /
\alpha_x) \eta]+(T/T_R) [ g_pF/(1+|\psi|^2)-\gamma]$,
 and $F=F(\xi,\eta)$ is the normalized
profile of the population inversion in the crystal. For a circular
Gaussian pump beam with spot size $w_p$ focused at the point
$(\rho_x,\rho_y)$ in the transverse plane, one has
$F(\xi,\eta)=\exp[-2(\xi L-\rho_x)^2/w_p^2-2(\eta
L-\rho_y)^2/w_p^2]$. Provided that the gain guiding is weaker than
the index guiding realized by the mirror tilt and reflection at the
edges, the output laser mode is expected to be given by a
two-dimensional truncated Airy beam, with its main lobe mostly
overlapped with the Gaussian pump beam. As an example, Fig.2 shows
the numerically-computed two-dimensional distributions of the most
unstable laser mode in a Nd:YVO$_4$ microchip laser, with a
transition from a Gaussian-like to a two-dimensional truncated Airy
profiles, for decreasing values of the logarithmic cavity losses
$\gamma$ and for $l=400 \; \mu$m, $\alpha_x=\alpha_y=0.2 \; {\rm
mrad}$, $\rho_x=\rho_y= 184 \; \mu$m, $w_p= 23 \; \mu$m.\\
In conclusion, it has been shown that microchip lasers with slightly
tilted facets can emit transverse output beams in the form of
truncated Airy beams. Such a result could be of interest for a
direct generation of Airy beams from an
ultracompact microchip device.\\
Work supported by the italian MIUR (Grant No. PRIN-2008-YCAAK).\\


\begin{thebibliography}{99}


%
%
%
%
%
%
%
%
%
%
%
%
%
%
%
%
%
%
%
%
%
%
%
%




\bibitem{m1}
J.J. Zayhowski and A. Mooradian, Opt. Lett. {\bf 14}, 24 (1990)

\bibitem{m2}
J.J. Zayhowski, Laser Focus World {\bf 35},  129 (1999).

\bibitem{m3}
P. Laporta, S. Taccheo, S. Longhi, O. Svelto, and G. Sacchi, Opt.
Lett. {\bf 18}, 1232 (1993).

\bibitem{m4}
G. K. Harkness and W. J. Firth, J. Mod. Opt. {\bf 39}, 2023 (1992).

\bibitem{m5}
T. Y. Fan, Opt. Lett. {\bf 19}, 554 (1994).

\bibitem{m6}
S. Longhi, J. Opt. Soc. Am. B {\bf 11}, 1098 (1994).

\bibitem{m7}
S. Longhi, G. Cerullo, S. Taccheo, V. Magni, and P. Laporta, Appl.
Phys. Lett. {\bf 65}, 3042 (1994).

\bibitem{m8}
A.J. Kemp, R.S. Conroy, G.J. Friel, and B.D. Sinclair, IEEE J.
Quant. Electron. {\bf 35}, 675 (1999).

\bibitem{m9}
C. Serrat, M.P. van Exter, N.J. van Druten, and J.P. Woerdman, IEEE
J. Quant. Electron. {\bf 35}, 1314 (1999).

\bibitem{a1}
G.A. Siviloglou, and D. N. Christodoulides, Opt. Lett. {\bf 32}, 979
(2007).

\bibitem{a2}
G.A. Siviloglou, J. Brokly, A. Dogariu, and D.N. Christodoulides,
Phys. Rev. Lett. {\bf 99}, 213901 (2007).

\bibitem{a3}
M. Mazilu, K. Dholakia, and J. Baumgart, Nature Photon. {\bf 2}, 675
(2008).

\bibitem{a4}
M.A. Bandres, Opt. Lett. {\bf 33}, 1678 (2008).

\bibitem{a5}
J. Broky, G.A. Siviloglou, A. Dogariu, and D.N. Christodoulides,
Opt. Express {\bf 16}, 12880 (2008).

\bibitem{a6}
P. Polynkin, M. Kolesik, J.V. Moloney, G.A. Siviloglou, and D.N.
Christodoulides, Science {\bf 324}, 229 (2009).

\bibitem{a7}
T. Ellenbogen, N. Voloch-Bloch, A. Ganany-Padowicz, and  A. Arie,
Nature Photon. {\bf 3}, 395 (2009).

\bibitem{a8}
I. Dolev, T. Ellenbogen, and A. Arie, Opt. Lett. {\bf 35}, 1581
(2010).

\bibitem{Svelto}
O. Svelto, {\it Principles of Lasers} (Fourth Edition, Springer,
1998).

\bibitem{FoxLie}
A. O. Fox and T. Li, Bell Syst. Tech. J. {\bf 40}, 453 (1961).

\bibitem{DellaValle09}
G. Della Valle, M. Savoini, M. Ornigotti, P. Laporta, V. Foglietti,
M. Finazzi, L. Duo, and S. Longhi, Phys. Rev. Lett. {\bf 102},
180402 (2009).

\bibitem{note1}
L.W. Casperson, J. Opt. Soc. Am. {\bf 66}, 1373 (1976).

\bibitem{note2}
 S. Longhi, Opt. Lett. {\bf 25}, 811 (2000)].



\end{thebibliography}
\end{document}